\documentclass[12pt]{article}
\usepackage[preprint, nonatbib]{neurips_2019} 

\usepackage[T1]{fontenc}
\usepackage[utf8]{inputenc}

\usepackage{tgtermes}
\usepackage{amsmath}
\usepackage{bbm}
\usepackage{scalefnt,letltxmacro}
\LetLtxMacro{\oldtextsc}{\textsc}
\renewcommand{\textsc}[1]{\oldtextsc{\scalefont{1.10}#1}}
\usepackage[scaled=0.92]{PTSans}
\usepackage{inconsolata}
\usepackage{mathbbol}

\usepackage[usenames,dvipsnames]{xcolor}
\definecolor{shadecolor}{gray}{0.9}

\usepackage[english]{babel}
\usepackage{afterpage}
\usepackage{framed}
\usepackage{nicefrac}

{\endMakeFramed}

\usepackage{booktabs}
\usepackage{multirow}
\usepackage{longtable}
\usepackage{etoolbox,siunitx}
\robustify\bfseries
\usepackage{bibunits}

\usepackage[algoruled,algo2e,linesnumbered]{algorithm2e}
\usepackage{listings}
\usepackage{fancyvrb}
\fvset{fontsize=\normalsize}
\usepackage{algorithm}
\usepackage{algorithmic}

\usepackage[colorlinks=true, linktoc=all, linkcolor=blue, urlcolor=blue, citecolor=blue]{hyperref}

\usepackage[nameinlink]{cleveref}

\usepackage[acronym,smallcaps,nowarn]{glossaries}
\glsdisablehyper{}

\usepackage{listings}
\lstdefinestyle{mystyle}{
    commentstyle=\color{OliveGreen},
    numberstyle=\tiny\color{black!60},
    stringstyle=\color{BrickRed},
    basicstyle=\ttfamily\scriptsize,
    breakatwhitespace=false,
    breaklines=true,
    captionpos=b,
    keepspaces=true,
    numbers=none,
    numbersep=5pt,
    showspaces=false,
    showstringspaces=false,
    showtabs=false,
    tabsize=2
}
\newcommand{\be}{\begin{equation}}
\newcommand{\ee}{\end{equation}} 
\newcommand{\bea}{\begin{eqnarray}}
\newcommand{\eea}{\end{eqnarray}}

\usepackage{graphicx}
\usepackage[labelfont=bf]{caption}
\usepackage{sidecap}
\sidecaptionvpos{figure}{t}

\newcommand{\tildeq}{\tilde{=}}

\usepackage[charter]{mathdesign}  

\begin{document}

\title{Sampling on networks: estimating spectral centrality measures and their impact in evaluating other relevant network measures.}

\author{
  Nicol\`{o} Ruggeri\\
  Max Planck Institute \\
  for Intelligent Systems,\\
  T\"ubingen\\
    \And
  Caterina De Bacco\\
  Max Planck Institute \\
  for Intelligent Systems,\\
    T\"ubingen\\
}

\maketitle

\begin{abstract}
  We perform an extensive analysis of how sampling impacts the estimate of several relevant network measures.  
  In particular, we focus on how a sampling strategy optimized to recover a particular spectral centrality measure impacts other topological quantities.  
Our goal is on one hand to extend the analysis of the behavior of \textit{TCEC} \cite{tcec}, a theoretically-grounded sampling method for eigenvector centrality estimation.
    On the other hand, to demonstrate more broadly how sampling can impact the estimation of relevant network properties like  centrality measures different than the one aimed at optimizing, community structure and node attribute distribution. 
    Finally, we adapt the theoretical framework behind \textit{TCEC} for the case of PageRank centrality and propose a sampling algorithm aimed at optimizing  its estimation. We show that, while the theoretical derivation can be suitably adapted to cover this case, the resulting algorithm suffers of a high  computational complexity that requires further approximations compared to the eigenvector centrality case. 
\end{abstract}

\section*{Introduction} \label{introduction}
When investigating real-world network datasets we often do not have access to the entire network information. This is the case of large datasets, having limited storage capacity or limited resources during the data collection phase. Nevertheless, this should not prevent practitioners from analyzing an available network sample. In fact, evaluating network properties while accessing only a smaller sample is a relevant problem in various fields, ranging from modeling dynamical processes \cite{de2010does,sadikov2011correcting}, network statistics estimation \cite{leskovec2006sampling}, data compression \cite{adler2001towards} and survey design \cite{frank2005network}. Imagining that one could design the sampling scheme for data collection, then this should be done wisely, as this biases the estimates of the network properties aimed at investigating \cite{han2005effect,lee2006statistical,kossinets2006effects}. The goal should be to design a sampling protocol that not only preserves the relevant network properties of the entire topology inside the sample, but that can be implemented efficiently.   
Most sampling strategies found in the literature \cite{leskovec2006sampling} are empirically-driven and lack of theoretical groundings. 
Recently, \textit{TCEC} \cite{tcec}, a sampling algorithm to approximate in-sample eigenvector centrality \cite{bonacich1972factoring}, whose main features are being theoretically grounded and computationally scalable, has been proposed. \textit{TCEC} aims at preserving the relative eigenvector centrality ranking of nodes inside the sample. This is a centrality measure used in many disciplines to characterize the importance of nodes. However, this might not be the only  property of interest when studying a network. The question is then how a sampling method, optimized to retrieve one particular property, performs in estimating other network-related measures. In this work we address this question by performing an extensive analysis of the behavior of \textit{TCEC} in recovering several relevant network  properties by means of empirical results on real networks. In particular, we focus on estimating various centrality measures which have a very different characterization from eigenvector centrality and do not come from spectral methods. Then we investigate how community structure and covariate information are affected by the sampling. We compare performance with other sampling strategies. Finally, we discuss what are the challenges preventing a trivial extension of \textit{TCEC} on PageRank \cite{brin1998anatomy}  score.

\subsection*{Related work}
 A large part of the scientific literature aiming at investigating sampling strategies on networks is based on empirical approaches \cite{blagus2017empirical,costenbader2003stability} and focus on recovering standard topological properties like degree distribution, diameter or clustering coefficient \cite{leskovec2006sampling,morstatter2013sample,stutzbach2009unbiased,hubler2008metropolis,stumpf2005sampling,ganguly2018estimation,antunes2018sampling}. To the best of our knowledge, \textit{TCEC} sampling \cite{tcec} is one of the first theoretical attempts in estimating eigenvalue centrality, which goes beyond heuristics or empirical reasoning. A closely related problem is that of estimating eigenvector centrality without observing any edge but only signals on nodes \cite{roddenberry2019blind}. A different but related research direction is to question the stability of centrality measures under perturbations \cite{segarra2015stability,han2016analysis,murai2019sensitivity}. 
 In the case of PageRank score, and more recently for Katz centrality as well \cite{lin2019sampling}, the focus of similar lines of research is based on the different objective of estimating single nodes' scores or approximating the external information missing for reliable within-sample estimation \cite{sakakura2014improved,chen2004local, davis2006estimating}, rather than estimating the relative ranking of nodes within a sample as we do here. Finally, focusing on temporal networks, \cite{shao2017inferring} propose a centrality measure suitable for this case and a method for its estimation using the network dynamics.

\section*{\textit{TCEC}: sampling for eigenvector centrality estimation}
\begin{figure}[t]
\begin{centering}
\includegraphics[scale=0.2]{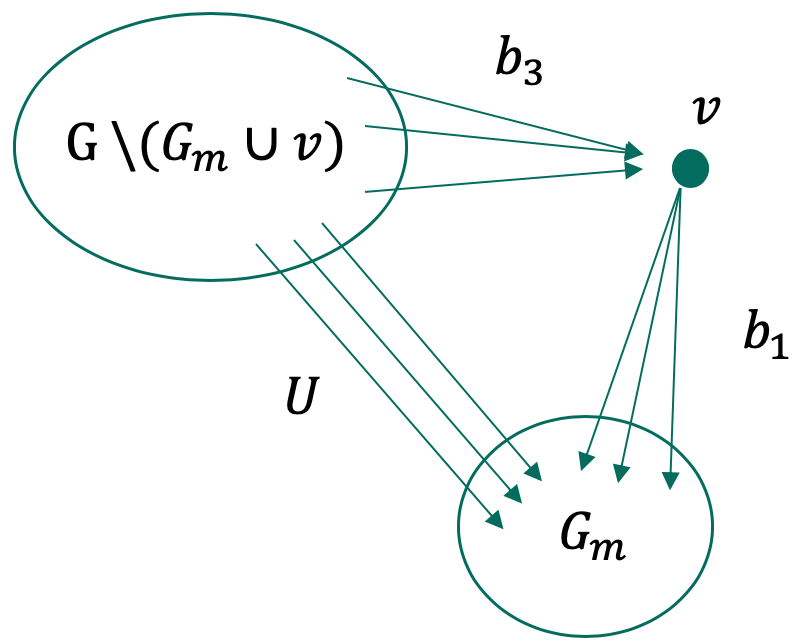}
\caption{\textit{TCEC} sampling visual representation. Consider a candidate node $v$ to be added to the current sample $G_m$. The algorithm considers:  the outgoing connections $b_1$ towards the sample, the incoming connections $b_3$ from the non-sampled nodes  and $U$, the remaining edges incoming towards the sample.}
\label{fig:sampling}
\end{centering}
\end{figure}

In this section we introduce the formalism and explain the main ideas behind the Theoretical Criterion for Eigenvector Centrality (\textit{TCEC}) sampling algorithm \cite{tcec}.
This method uses mathematical formalism from spectral approximation theory to approximate the eigenvector centralities of nodes in a subsample with their values in the whole graph. Consider a graph $G=( \mathcal{V}, \mathcal{E})$ where $ \mathcal{V}$ is the set of nodes and $ \mathcal{E}$ the set of edges; denote $A$ its adjacency matrix with entries $A_{ij} \in \mathcal{R}_{\geq 0}$ the weight of an edge from $i$ to $j$. Sampling a network can be defined as the problem of selecting a \textit{principal submatrix} $A_{m}^{'}$ of size $m\leq |\mathcal{V}|$ induced by a subset of nodes $\mathcal{I}\subseteq \mathcal{V}$. The subsampled network is denoted as $G_{m}=( \mathcal{I}, \mathcal{E}_{m})$, and $\mathcal{E}_{m} \subseteq \mathcal{E}$ is the set of edges in the subsample. In general, there can be several choices for selecting $G_{m}$. They should depend on the quantities aimed at preserving when sampling.  \textit{TCEC} selects $G_{m}$ in order to minimize the sin distance $\sin(\mu_{m},\tilde{\mu})$ between the eigenvector centrality $\tilde{\mu} \in \mathcal{R}^{m}$ in the subsample and the one on the same nodes, but calculated from the whole graph $\mu_{m} \in \mathcal{R}^{m}$; $\mu_m$ is a vector built from the whole-graph eigenvector centrality $\mu \in \mathcal{R}^{V}$, when selecting only the $m$ entries corresponding to nodes in the subsample. Accessing $\sin(\mu_{m},\tilde{\mu})$ without the knowledge of the whole graph is not possible. However, given that eigenvector centrality is a spectral method, i.e. is based on evaluating eigenvectors and eigenvalues, \textit{TCEC} uses projection methods for spectral approximation to propose a bound on that distance and relate it to network-related quantities. \\
This results in an algorithmic implementation of a sampling procedure that aims at minimizing that bound. Referring to \cite{tcec} for details, the algorithm briefly works as follows. Starting from an initial small random sample, it selects nodes in an \textit{online} fashion: it adds to the current sample $\mathcal{I}$ of size $k$ one node at a time by selecting the best node from the set of non-sampled nodes $j\in \mathcal{V} \setminus \mathcal{I}$. The best candidate node $j$ is the one that maximizes the following quantity made of network-related quantities:
\begin{equation}\label{eqn:theor crit alpha}
\null \hspace{2cm} (1 -\alpha) \left(  || b_1 ||_2^2 + || b_1^T U ||_2^2 - || b_3 ||_2^2  \right) + \alpha \,d_{in}^{G_k}(j) \quad,
\end{equation}
where $b_1\in \mathbb{R}^{k-1}$ are the edges pointing from $j$ to the nodes already in the subsample, $b_2 \in \mathbb{R}$ is the entry corresponding to $j$, $b_3 \in \mathbb{R}^{n-k+1}$ are edges from nodes outside the sample towards $j$, $U \in \mathbb{R}^{k-1, n- k +1}$ are the edges from nodes outside the sample towards nodes in it, $j$ excluded; $d_{in}^{G_{k}}(j)$ is the (weighted) in-degree of node $j$ calculated considering only the incoming edges from nodes that are in the sample;
$\alpha \in [0,1]$ is an hyperparameter that can be tuned empirically. We present a diagram of the quantities involved in Fig. \ref{fig:sampling}.

\section*{Empirical studies}\label{sec:empirical}
We study the impact of sampling a network with \textit{TCEC} on several relevant network properties different form eigenvector centrality. Namely, we investigate: i) the distribution of the sampled nodes in terms of non-spectral centrality measures as \textit{in-degree}, betweenness centrality and SpringRank \cite{de2018physical}; ii) the relationship between community structure and sampled nodes; iii) the preservation of the distribution of node attributes in the sampled network. For all these tasks, we compare with uniform random walk sampling (RW), as this is the mainstream choice for many sampling scenarios, due to its favorable statistical and computational properties \cite{gjoka2010walking}; it has also shown better performance in recovering eigenvector centrality than all other state-of-the-art algorithms analyzed against \textit{TCEC} \cite{tcec}. In addition, in the absence of a best sampling protocol that works for all applications, we further compare with a third algorithm, chosen differently according to the task at hand.

\paragraph*{Implementation details}  While we refer to \cite{tcec} for the detailed definitions of the parameters needed in the algorithmic implementation, we provide a summary of their values used in our experiments in the Appendix B; we used the open-source implementation of \textit{TCEC} available online\footnote{\url{https://github.com/cdebacco/tcec_sampling}}. 

\paragraph*{Non-spectral centrality measures behavior} We analyzed the performance of \textit{TCEC} in estimating non-spectral centrality measures in real world datasets: 
the \textit{Epinions dataset}\footnote{\url{https://snap.stanford.edu/data/soc-Epinions1.html}} \cite{takac2012data}, a who-trusts-whom dataset based on the review site \textit{Epinions.com};
the \textit{Slashdot dataset}\footnote{\url{https://snap.stanford.edu/data/soc-Slashdot0811.html}} \cite{leskovec2009community}, a social network based on the reviews website \textit{Slashdot.org} community;
the \textit{Stanford network} \footnote{\url{https://snap.stanford.edu/data/web-Stanford.html}} \cite{leskovec2009community}, a network of hyperlinks of  the \textit{stanford.edu} domain. We considered here only directed networks as this is the relevant case for the centrality measures we are considering.

\begin{figure}[hptb]
\hspace*{0.7cm}
\includegraphics[scale=0.32]{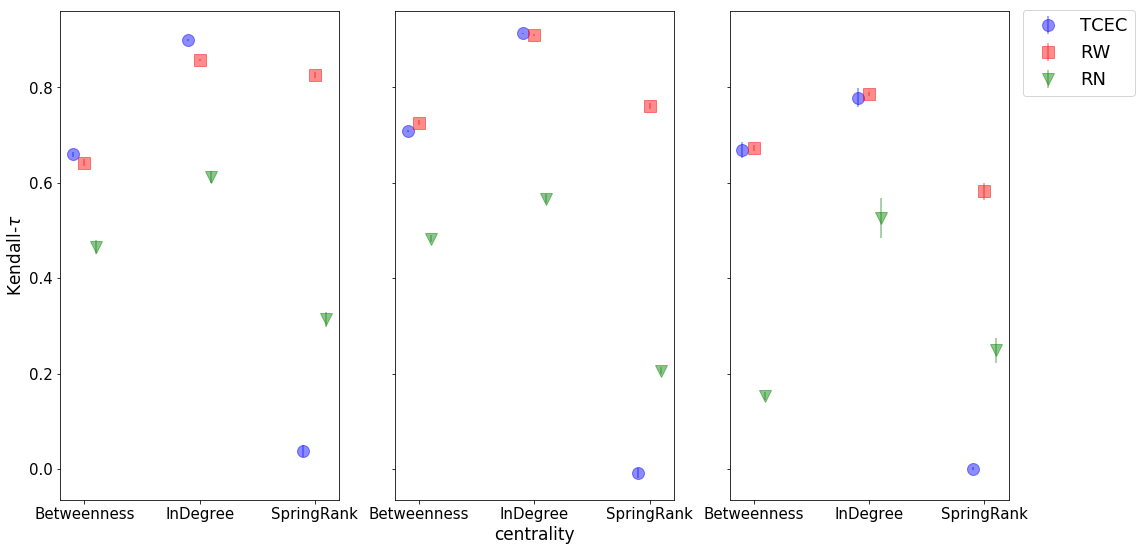}
\caption{\small
Approximation of centrality measures with \textit{TCEC}, RW and uniform sampling for the three different datasets, Epinions (left), Slashdot (center) and Stanford.edu (right). Scatterplots represent average results over 10 runs on the three datasets, with standard deviation indicated by vertical lines. 
}
\label{fig:centrality}
\end{figure}

We compared with RW and uniform sampling on nodes (RN), since this is a commonly used sampling criterion for generic tasks \cite{leskovec2006sampling,ahmed2012network, wagner2017sampling}. We consider three different centrality measures: i) in-degree centrality, which corresponds to the in-degree of a node; ii) betweenness centrality, a measure that captures the importance of a node in terms of the number of shortest paths that need to pass through it in order to traverse the network; iii) SpringRank \cite{de2018physical}, a physics-inspired probabilistic method to rank nodes from directed interactions which yields rank distributions relatively different than that of spectral measures, like eigenvector centrality. Together, these three provide a diverse set of methods to characterize a node's importance. Importantly, none of this is based on spectral methods, which represents the theoretical grounding behind \textit{TCEC}. \\
As we show in Fig. \ref{fig:centrality}, both betweenness and in-degree centrality are well approximated by RW and \textit{TCEC} on all datasets. The SpringRank score is the most discriminative between sampling algorithms. In this case RW succeeds in retrieving significant interactions to well approximate rankings, while RN performs poorly and \textit{TCEC} yield Kendall-$\tau$ correlation close to 0. SpringRank aims at inferring hidden hierarchies of nodes from directed pairwise interactions. We argue that \textit{TCEC} performs poorly in recovering SpringRank values similar to the ones in the whole graph because may cut relevant information for this task: being biased towards nodes with many connections towards the sample but few incoming connections from outside (see Fig. \ref{fig:sampling}), can change fundamentally the structure of directed pairwise interactions (i.e. edges inside the sample) at the core of SpringRank. Instead, RN cuts discriminative edges by not taking the topology into account at sampling time, and therefore achieves poor performances in recovering any edge-based centrality measure.

\begin{figure}[hptb]
\includegraphics[scale=0.165]{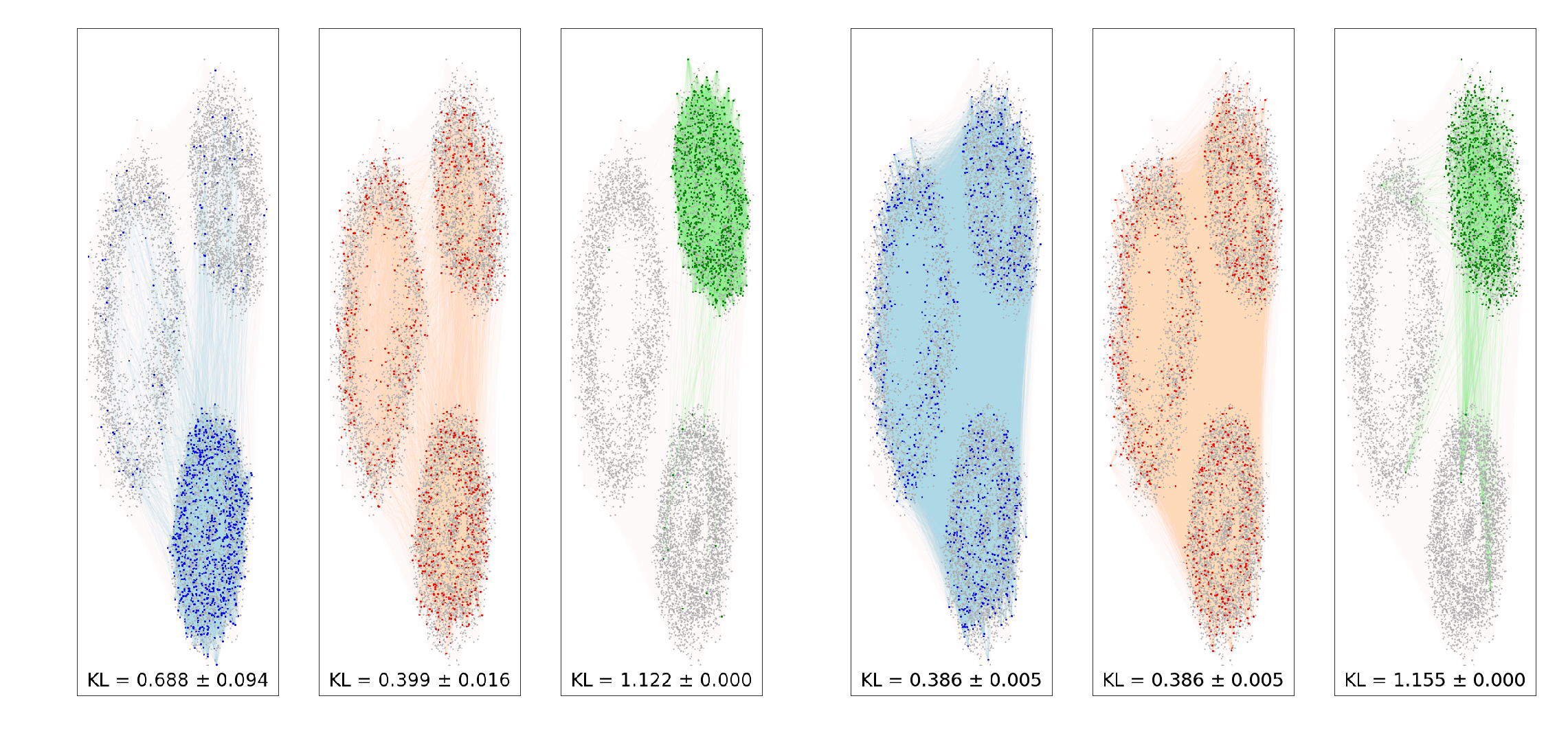}
\caption{
\small Community structure and sampling. We show an example of sampling two synthetic SBM networks one (left) assortative and one (right) disassortative. Sampled nodes and edges are colored in blue for \textit{TCEC} (left), red for uniform random walk  (center) and green for expansion sampling (right). The KL-divergence averages and standard deviations are computed over 10 different rounds of sampling $10\%$ of all the nodes. 
} 
\label{fig:SBM}  
\end{figure}

\paragraph*{Community structure preservation} We investigate how the sampling algorithms impact an underlying network community structure. To this end, we study the distribution of the community memberships of sampled nodes in synthetic networks generated with Stochastic Block Model (SBM) \cite{holland1983stochastic} of size $N=10000$ nodes divided in 3 communities, where we sample $10\%$ of the nodes.  Sampling protocols can be sensitive to the topological structure of the network (assortative or homophilic, disassortative or heterophilic) and to the balance of group sizes \cite{wagner2017sampling}. These can all impact how the different groups are represented in the sample and other factors such as individuals' perception biases \cite{lee2019homophily}. We thus run tests on both types of structures and using various levels of balance for the communities. Specifically, we consider 
i) balanced assortative networks: two groups of 3000 nodes and one of 4000, within-block probability of connection $p_{in}=0.05$ and between-blocks  $p_{out}=0.005$; ii) unbalanced assortative networks: groups of sizes 1000, 3000 and 6000 respectively, same $p_{in}$ and $p_{out}$ as in i); iii) balanced disassortative networks: same group division as in i) but within-block probability of connection $p_{in}=0.005$ and between-blocks  $p_{out}=0.05$. 
We compare \textit{TCEC} with RW, which was shown to be robust in representing groups in the sample \cite{wagner2017sampling} and expansion sampling \cite{maiya2010sampling}, since it has been explicitly built to sample community structure. All algorithms start sampling from a node belonging to the group of smallest size. We observe two qualitatively different trends in the way nodes are chosen.
Random walk yields samples of nodes more homogeneously distributed across communities, in all network structures. \textit{TCEC}, instead, tends to select nodes within the block where it has been initialized. A possible explanation for this behavior is given by the peculiar form of the \textit{TCEC} score of Eq. (\ref{eqn:theor crit alpha}). This in fact tends to select nodes with a large $|| b_1 ||_2$ and small $|| b_3 ||_2$, i.e. many connections towards the sample and few connections from outside the sample. A likely choice is to then select nodes within the same community, where this combination holds. Notice that the nodes outside of the main sampled community can be attributed to the random walk initialization before the main \textit{TCEC} routine. Finally, expansion sampling remains confined in a single block, as it is a  deterministic algorithm. This can be computationally prohibitive for deployment with larger sample sizes.
Results are presented in Fig. \ref{fig:SBM}, where we also report the KL-divergence \cite{kullback1951information} between the communities distribution in the sample and the whole network for all sampling algorithms.
The KL-divergence is a measure of discrepancy between probability distributions, which is 0 if they perfectly overlap, and it gets larger as the difference between them grows. Thus, higher values signal higher discrepancy between the in-sample block distribution and the one calculated on the entire network. This can be observed graphically in Fig. \ref{fig:SBM} (left) for the assortative homogeneous structure i). Here the higher KL divergence is due to a more pronounced clustering of sampled nodes in one single block. The nodes selected by RW are more scattered around different blocks, while \textit{TCEC} tends to select nodes within a single block and expansion sampling is completely confined to the initial one. Similar results hold for case ii), as defined above,  and are presented in Appendix D. 
For the disassortative structure iii), however,  results differ. In this case, \textit{TCEC} and RW tend to explore the network in a similar manner. A lower KL-divergence from the ground truth signals the fact that blocks are sampled more uniformly. While for RW this phenomenon is explained by the stochasticity of the neighbourhood exploration, for TCEC it is caused by the way the algorithm works in selecting candidate nodes with high out-degrees towards the sample but small in-degrees from outside of it, as shown in Fig. \ref{fig:sampling}. In disassortative networks these likely candidates belong to different communities, thus the more homogeneous exploration.  Expansion sampling is still confined inside the starting block as in the previous case.

\paragraph*{Node attribute preservation} Another relevant question is whether node attributes are affected by the way the network is sampled. This is particularly important in cases where extra information is known, along with the network's topological structure. For instance, in relational classification,   network information is exploited to label individuals (e.g. recovering nodes' attributes); classification performance can significantly change based on the sampling protocol adopted \cite{ahmed2012network,espin2018towards}.

\begin{figure}[htb]
\hspace*{0.7cm} 
\includegraphics[scale=0.36]{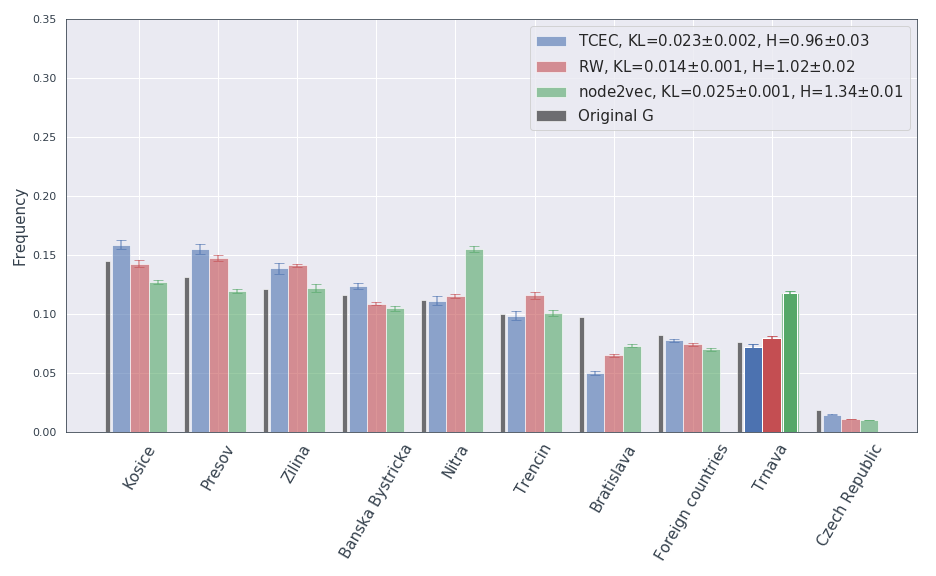}
\caption{\small
Covariates distribution and sampling. We consider the Pokec social network dataset ($\approx 1.3 \cdot 10^6$ nodes and $\approx 2.9 \cdot 10^7$ edges) with a sample fraction of $10\%$. The columns with bolded colors represent the region where we started the sampling from. Numbers inside the legend are average and standard deviations over 10 runs of $KL(p_{G}||p_{G_{m}})$, where $p_{G}$ is the empirical frequency of the ten regions in the original graph, similarly for $p_{G_{m}}$ in the sample. $H$ represents the mean and standard deviation over the same runs for the $H_{G_m}(s) / H_G (s)$ entropy ratio. We plot here an example of the relative frequency of nodes for the ten regions in which nodes are divided. The distribution on the whole network  is the black vertical line. Vertical lines on top of bars represent standard deviations across 10 runs of sampling. Notice three different behaviors: RW obtains an in-sample attribute distribution similar to the one on the whole graph. \textit{TCEC} has a higher difference in KL, followed by \textit{node2vec}. On average, the former two are not biased by the starting region, as it is instead the case for \textit{node2vec}. This can also be observed quantitatively by a higher $H_{G_m}(s) / H_G (s)$ ratio. 
}
\label{fig:pokec}
\end{figure}

In general, when performing statistical tests on sampled networks' covariates, we work under the assumption that their distribution is similar to that of the original network. However, this assumption is not necessarily fulfilled when performing arbitrary sampling. Notice that this is a related but different problem than the one above of community structure preservation. In that case, we were explicitly imposing that communities are correlated with network structure. In case of attributes, we can only assume that, but this may not be valid depending on the real dataset at hand.
We test this behavior by studying the \textit{Pokec dataset}\footnote{\url{https://snap.stanford.edu/data/so-Pokec.html}} \cite{takac2012data}. This is a social network representing connections between people in form of online friendships. In addition, the dataset contains extra covariate information on nodes, i.e. attributes about the individuals. In our case we focus on one of them, the geo-localization of users in one of the ten regions (the eight Slovakian regions, Czech Republic and one label for all other foreign countries) where the social network is based. We compare the distribution of this covariate in the full network with that on the nodes sampled by random walk,  \textit{TCEC} and \textit{node2vec} \cite{grover2016node2vec}, with exploration parameters $p=2$, $q=0.5$, i.e. depth-first oriented search. The choice of \textit{node2vec}  is motivated by its frequent implementation for node embedding tasks. As node embeddings are often used for regression or classification tasks, along with network covariates, it is thus relevant for our task here.
We run the algorithms starting from seed nodes within different regions, as the choice of the initial sample
of labeled seed nodes can impact the final in-sample attribute distribution \cite{wagner2017sampling}. As before, we measure KL-divergence between the empirical attribute distribution on the entire network against that found within the sample. A graphical representation of one example of the results is given in Fig. \ref{fig:pokec}. We notice different behaviors for the various sampling methods.
While all algorithms recover a covariate distribution close the ground truth, slightly better performances are achieved, in order, from RW, \textit{TCEC} and \textit{node2vec}, with average KL values ranging from 0.01 to 0.04 respectively. However, a peculiar trend can be observed in relation to the starting region. In fact, the final sample is biased towards over representing the seed region for \textit{node2vec}, as opposed to a comparable homogeneity obtained by \textit{TCEC} and RW. This is a subtle result, as this over representation is not shown by the KL values. 
Instead, it can be measured by the entropy ratio $H_{G_{m}}(s)/H_{G}(s)$ between the entropy $H_{G_{m}}(s)=-p_{G_{m}}(s)\log p_{G_{m}}(s)-(1-p_{G_{m}}(s))\log(1-p_{G_{m}}(s)) $ of a binary random variable representing whether a node in the sample belongs to the seed region $s$ or not, over $H_{G}(s)$, the same quantity  but calculated over all nodes in the graph. In words, this measures the discrepancy of the frequency of the particular attribute corresponding to the seed region between in-sample nodes and the whole network. Values close to 1 denote high similarity, greater than 1 means over representation and less than one under representation of a particular attribute. In all but two starting regions, \textit{node2vec} has a significantly high entropy ratio: for various seed regions this is higher than 1.19 whereas the maximum values obtained by \textit{TCEC} and RW are both less than 1.12. Quantitatively, this shows the magnitude of the  over representation in the sample induced by \textit{node2vec};  instead, \textit{TCEC} and RW do not yield any significant bias towards the starting region.  An example of this behavior is plotted in Fig. \ref{fig:pokec}, all the other starting regions are given in Appendix C.

\section*{Sampling for PageRank estimation}
In this section we discuss the challenges preventing an effective extension of the theoretical framework behind \textit{TCEC} to PageRank score (PR) \cite{brin1998anatomy} , i.e. a method for sampling networks theoretically grounded on the same ideas, but aiming at better approximating PageRank, rather than eigenvector centrality. In fact, arguably counterintuitively, there is no trivial generalization of \textit{TCEC} for PageRank. Instead, it is necessary to make further assumptions that result in an algorithmic scheme that is equivalent to \textit{TCEC} in practice, from our empirical observations. Here we explain the main challenges and refer to the Appendix for detailed derivations of how to address them.
PageRank considers a different adjacency matrix $A_{PR}$, which is strongly connected (as the network is complete) and stochastic (the rows are normalized to $1$). This is built from the original $A$. Both these features, not present for the eigenvector centrality case, are the cause of the additional complexity of sampling for PageRank.  The PR score is defined as the eigenvector centrality computed on $A_{PR}$. At a first glance, this may lead to a straightforward generalization of \textit{TCEC} sampling by simply applying the algorithm to $A_{PR}$. However, this simple scheme hinders in fact one main challenge, which makes this generalization theoretically non trivial.
\textit{TCEC} yields the matrix $A_{G_{m}}$ (the adjacency of the sampled network $G_m$ ), which \textit{is} a submatrix of the original $A$; having a submatrix is a requirement for the validity of the sin distance bound at the core of \textit{TCEC}. Instead, in the case of PageRank, the matrix of the sampled network $A_{PR, G_{m}}$  is \textit{not} a submatrix of $A_{PR}$; this is because $A_{PR}$ is a stochastic matrix, which requires knowing the degree of each node in advance to normalize each row. This information is in general not known a priori. We fixed this problem introducing an approximation (see Appendix) which allows to use the theoretical criterion of Eq. (\ref{eqn:theor crit alpha}) in this case as well. However, we still face a computational challenge. Due to the nature of PageRank, which allows jumps to non-neighboring nodes, albeit with low probability, the networks behind $A_{PR}$ and  $A_{PR, G_m}$ are both complete. This results in a much higher computational cost of the sampling algorithm. Even though we proposed ways to fix this issue as well (see Appendix)  and thus combined these two considerations into an efficient algorithmic implementation (which we refer to as \textit{TCPR}) analogous to \textit{TCEC}, empirical results for this are poor. In practice, \textit{TCEC} performs better in recovering the PR scores of nodes in the sample.

\paragraph*{\textit{TCEC} vs \textit{TCPR} for PageRank approximation} We compare the approximation of the PageRank score as obtained on samples from random walk, \textit{TCEC} and \textit{TCPR}, via Kendall-$\tau$ correlation \cite{kendall1948rank} with the true score, which were assumed to be available in these experiments. A higher correlation signals a better recovery of the relative ranks between nodes. We do so on the Epinions,  Internet Topology, Slashdot and Stanford network. The \textit{Internet Topology dataset}\footnote{\url{http://irl.cs.ucla.edu/topology/}} \cite{zhang2005collecting}, represents the (undirected) Internet Autonomous Systems level topology. \\
For these experiments we set the \textit{TCEC} randomization probability to 0.5, to achieve better approximation scores (see appendix B). Figure \ref{fig:Pr experiment} shows a noticeable improvement of \textit{TCEC} in most of the networks, both as a function of the sampling ratio and compared to RW for in-sample PR ranking recovery. However, we do not observe such a pattern for \textit{TCPR}, which performs better than \textit{TCEC} only for few datasets and sample ratio combinations. As the theoretical groundings behind the two are similar, we argue that using the $L_{1}$-norm in \textit{TCPR} (see Appendix A), which is inherently less discriminative of the $L_{2}$-norm behind \textit{TCEC}, seems to affect this difference in performance. Another possible cause is the extra assumption of in-sample nodes' degrees linearly scaling with sample size. Large deviations from this assumption could sensibly impact the quality of the goodness criterion at hand. 

\begin{figure}[hptb]
\hspace*{-0.8cm}\includegraphics[scale=0.3]{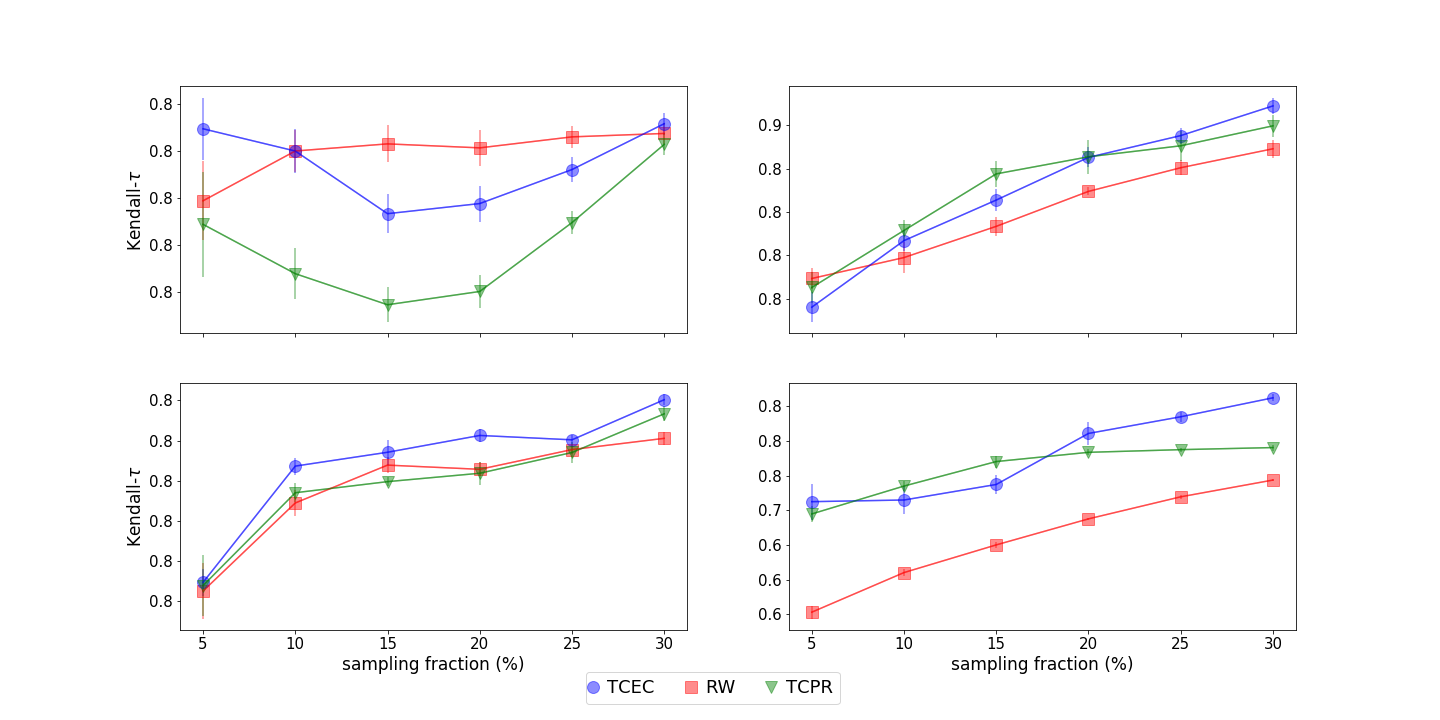}
\caption{\small
Results on the four datasets for PR score approximation, respectively Epinions (upper left), Internet Topology (upper right), Slashdot (lower left) and Stanford (lower right). While, as a general trend, \textit{TCEC} and \textit{TCPR} seem to perform in average better than random walk for PR score estimation, there is no clear separation between the former two. Standard deviations are computed on 10 runs of sampling.
}
\label{fig:Pr experiment}
\end{figure}

\section*{Conclusions}
Designing a sampling protocol when the whole-network information is not accessible is a task that has to be performed wisely. 
In fact, the choice of the sampling algorithm biases the analysis of relevant network quantities performed on the sample.
We investigated here the impact on various centrality measures, community structure and node attribute distribution that sampling techniques have. 
We studied in particular the performance of \textit{TCEC}, a theoretically grounded sampling method aimed at recovering eigenvector centrality
on such network properties within the sample and compared with other sampling approaches. The goal was to understand whether a sampling algorithm  optimized to preserve a specific global and spectral network measure, is indirectly preserving also other network quantities.
We empirically found that on various real networks \textit{TCEC} performs relatively differently than other sampling algorithms on the various tasks. In particular, while it performs better than uniform random walk in recovering PageRank values, i.e. a spectral measure, it yields uncorrelated rankings in terms of SpringRank, a non-spectral centrality measure which behaves qualitatively very differently than eigenvector centrality, signaling that \textit{TCEC} sampling might break the topological structure needed for SpringRank recovery. In addition, while RW yields community structure homogeneously distributed across blocks, \textit{TCEC} tends to select nodes inside the starting community, however partially reaching out to other blocks. Finally, studying a large online social network, it recovers in-sample attribute distributions close to the ones of the whole graph. It does not show any significant bias towards the seed region, as it is instead the case for \textit{node2vec}, which is over representing the starting regions.
We discussed possibilities of extending \textit{TCEC} to the case of PageRank and showed the challenges associated to this task and the remedies to them. However, the resulting algorithm performs comparably well to \textit{TCEC} on recovering PageRank values.
We focused here in showcasing the impact of sampling on three different relevant tasks that have broad relevance in network datasets. It would be interesting to extend a similar type of investigation to more specific network-related measures in concrete applications. For instance, understanding the mechanism why \textit{TCEC} gives SpringRank values almost uncorrelated to those on the original network would provide useful insights on how to break or preserve relevant structural network properties.

%
%

\section*{Acknowledgements}
     We are grateful to  Mirta Galesic for helpful discussions.  

\bibliographystyle{vancouver} 
\bibliography{bibliography}      

\section*{Appendix A. \textit{TCPR}: extending the model to PageRank score}
Before introducing the theory behind \textit{TCPR}, we begin with a review of the basic PageRank algorithm, introduce some notation and outline the main challenges and assumptions needed in the following derivations.

\paragraph*{Notation} 
Consider a nonnegative adjacency matrix $A \in \mathbb{R}^{V, V}$. Then we build a new adjacency matrix $A_{PR}$, called \textit{PageRank adjacency matrix}, defined as follows
\begin{equation}\label{eq: Apr matrix}
A_{PR} := \gamma P + \frac{(1 - \gamma)}{V} e e^T   \, ,
\end{equation} 
where $e$ is the vector of ones of length $V$, $\gamma \in [0, 1)$ and $P$ is defined as 
\begin{align}
P &= Q + \frac{e d^T}{V}   \label{eq: P matrix} \\
d &= e^T - e^T Q   \label{eq: d vector} \\
Q &= A \cdot  \text{diag} \left(\frac{1}{d_1}, \ldots, \frac{1}{d_V} \right)  \, ,    \label{eq: Q matrix}
\end{align}
with $d_j = \sum_i A_{ij}$ out-degree of node $j$ and with the convention that for node with zero out-degree,  named \textit{dangling nodes}, we take $1 / d_j = 0$.   \\
For all the matrices $P, Q, A_{PR}$ we can define two different quantities. Given a subset of nodes $\{1, \ldots, m \}$ of $V$ we have the principal submatrices $P_m, Q_m, A_{PR, m}$ relative to these nodes. But we can also compute the PageRank scores on the subgraph $G_m$. 
The matrices relative to $G_m$ are instead noted as $P_{G_m}, Q_{G_m}, A_{PR, G_m}$. Notice that for the original case of eigenvector centrality we had the correspondence $A_m = A_{G_m}$, we will simply refer to this as $A_m$. \\

\paragraph*{Challenges}
In general, as we sample, we only know $d_{in}^{G_{m}}(i)$ but may not have access to $d_{in}^{G}(i)$ ( in general $d_{in}^{G_{m}}(i) \leq d_{in}^{G}(i)$); this implies that the entries of $A_{PR, G_{m}}$ are different than the submatrix $A_{PR,m}$ of $A_{PR}$ induced by the nodes in $G_{m}$. We tackle this challenge by making an additional assumption: we assume that the degree $d_{in}^{G_{m}}(i) \approx \frac{m}{V}\, d_{in}(i) $, i.e. degrees of nodes in the sample scale linearly with the sample size $m$; this is a necessary approximation for linking the two otherwise different matrices $A_{PR,m}$ and $A_{PR, G_{m}}$ (which where instead equal for eigenvector centrality), its validity has been justified \cite{ganguly2018estimation} and thus we can use the theoretical criterion of Eq. (\ref{eqn:theor crit alpha}) in this case as well. 
This fixes a theoretical challenge, however, we now face a computational one. Due to the nature of PageRank, which allows jumps to non-neighboring nodes, albeit with low probability, the networks behind $A_{PR}$ and  $A_{PR, G_m}$ are both complete. This results in a much higher computational cost of the sampling algorithm. We reduce this by selecting candidate nodes to be added to the sample, in analogy with \textit{TCEC}, among the \textit{incoming} neighbors only, thus neglecting nodes that correspond to a non-zero entry of $A_{PR}$ but do not correspond to an actual edge. This has also the advantage of excluding dangling nodes (i.e. nodes with out-degree zero) from the sample. 
Combining these two considerations, we obtain a sampling criterion similar to the one employed in \textit{TCEC}; we name this \textit{TCPR} (Theoretical Criterion PageRank).

\paragraph*{Adapting the theory to PageRank}
While for "vanilla" eigenvector centrality the matrix $A$ was by hypothesis sparse, and therefore border exploration feasible, now the network represented by $A_{PR}$ is complete.
Border exploration, even if randomized by a level $p$, would be of cost $O(V)$. For \textit{TCEC} the choice was to choose all {\it incoming} neighbours in the sample. Here we can do the same, but {\it only choosing incoming neighbours from the original network} $A$. This because an incoming connection in 
$A$ has weight $\sigma(1/d_{out})$ in $A_{PR}$, while one due to the artificial edges in \eqref{eq: P matrix} and \eqref{eq: Apr matrix} have total weight $\sigma(\frac{k+1}{V})$, which is negligible for sample size $k << V$. \\
This is also in line with the observation that in many sampling scenarios we are not really able to pick nodes in the graph at random, but just explore neighbourhoods \cite{gjoka2010walking}. Additionally, by only considering incoming neighbours, which have out-degree necessarily greater than 0, we exclude all the dangling nodes form the final sample. \\

Notice that the theorem in \cite{tcec} was comparing the principal eigenvectors of a matrix $A$ and a principal submatrix $A_m$. In the case of PageRank, this is not applicable. In fact the matrix $Q$ from eq \eqref{eq: Q matrix} is normalized differently. In $A$, the rescaling is done on the full graph,
while in $A_m$ on the subgraph degrees. This means that $Q_m \neq Q_{G_m}$, and consequently $P_m \neq P_{G_m}$, $A_{PR} \neq A_{PR, m}$. This problem can be overcome by making a further assumption. In a pseudo-random choice of any subsample of size $m$, it reasonable to assume that nodes' degrees scale linearly, 
i.e. $N_{j, G_m} \approx \frac{m}{V} N_j $. By holding this approximation as valid, and recalling that there are no dangling nodes in the subgraph, it is straightforward to check that $A_{PR, G_m} = \frac{V}{m} A_{PR, m}$. In particular, the eigenvector centrality for sampled nodes is the same in the complete graph $G$ and the sampled one $G_m$, since $A_{PR, G_m}, A_{PR m}$ have the same eigenvectors. This overcomes the first issue of linking the PR score on $G_m$ and $G$, and we can simply sample nodes with the goodness criterion from \cite{tcec} on the page rank matrix $A_{PR}$. \\

We are left with the necessity of computing the goodness criterion efficiently.

\paragraph*{Efficient criterion computation}  
Suppose, without loss of generality,  that the sampled nodes are $\{1, \ldots, k  \}$ and the new node under evaluation $k+1$. Considering the PR adjacency matrix $A_{PR}$ the quantities involved in the theoretical criterion \eqref{eqn:theor crit alpha} are:
\begin{align}
	b_1 	&= \alpha P_{:, k+1} + \frac{1 - \alpha}{V}e  \label{eq: b1} \\
	b_3	&= \alpha P_{k+1, :} + \frac{1 - \alpha}{V}e  \label{eq: b3} \\
	U		&= \alpha P_{1:k, k+1:n} + \frac{1 - \alpha}{V} \mathbbm{1} \, , \label{eq: U}
\end{align}
where $e, \mathbbm{1}$ are respectively a vector and matrix of all ones, of correct dimensions. Moreover, we need $b_1^T U$. By explicit calculations:
\begin{align}\label{eq: b1_U}
b_1^T U		&= \alpha^2 P_{:, k+1}^T P_{1:k, k+1:n} + \frac{\alpha (1-\alpha)}{V} P_{:, k+1} \mathbbm{1}    \nonumber \\
				&\hspace{4mm} + \frac{\alpha (1-\alpha)}{V} e^T P_{1:k, k+1:n} + \left( \frac{1-\alpha}{V} \right)^2 e^T \mathbbm{1}  \, .
\end{align}

Implementing the computation of $b_1^T U$ in sparse arithmetics is not convenient, as it would anyway cost $O(k)$. Performing this increasingly costly operation for all (or some) of the nodes in the border at every new node sampled is not feasible. Here we optimise this computation explicitly. First, notice from equation \eqref{eq: b1_U} that
many terms are independent on the sample. Therefore we compute the $L_{1}$-norm (\cite{kamvar2003extrapolation}) for all the vectors \eqref{eq: b1}, \eqref{eq: b3}, \eqref{eq: b1_U}. In all the following computations we use the symbol $\tildeq$ to indicate equality up to an additive constant independent on the sampled node $k$. $a_{ij}$ stands for the element $i, j$ of $A$. 
\begin{itemize}
\item term $b_1$: 
	$$|| b_1 ||_1 \tildeq \alpha || P_{:, k+1} ||_1 = \frac{\alpha}{N_k} \sum_{j \in G_k} a_{jk} $$

\item term $b_3$:  
	$$|| b_3 ||_1 \tildeq \alpha || P_{k+1, :} || = \alpha\left( \sum_{\substack{i \notin G_k \cup \{k+1\} \\  i \text{ not dangling}  }}   \frac{a_{ki}}{N_i}  +  \sum_{\substack{i \notin G_k \cup \{k+1\} \\  i \text{ dangling}  }}  \frac{1}{n}  \right) \tildeq \alpha \sum_{\substack{i \notin G_k \cup \{k+1\} \\  i \text{ not dangling}  }}   \frac{a_{ki}}{N_i}  $$
where the last equality is justified by the fact that, since $k+1$ cannot be dangling, $\{i \notin G_k \cup \{k+1\} \, : \,  i \text{ dangling}  \} = \{i \notin G_k \, : \,  i \text{ dangling}  \}$, which is independent on $k+1$.

\item term $b_1^T U$. For this we need to split the computation in three, since from equation \eqref{eq: b1_U}: 
$$|| b_1^T U || \tildeq \alpha^2 ||P_{:, k+1}^T P_{1:k, k+1:n}||_1 + \frac{\alpha (1-\alpha)}{n} ||P_{:, k+1} \mathbbm{1}||_1 + \frac{\alpha (1-\alpha)}{n} ||e^T P_{1:k, k+1:n}||_1$$
For every node $j \in G_m$ define $\delta_{j} := \sum_{i \notin G_k \cup \{ k+1 \} } P_{ij}$ (which also depends on the subsample $G_k$ and the new proposal node $k+1$, we omit the dependence in the notation). Then:
\begin{align}
 ||P_{:, k+1}^T P_{1:k, k+1:n}||_1	&= \frac{1}{N_k} \sum_{j \in G_k} a_{jk} \sum_{i \notin G_k \cup \{ k+1 \}} P_{ij} =  \frac{1}{N_k} \sum_{j \in G_k} a_{jk} \delta_{j}    \label{eq: efficient expression}\\
 ||P_{:, k+1} \mathbbm{1}||_1 	&= \sum_{\i \notin G_k \cup \{ k+1 \} } \sum_{j \in G_k} P_{ji}   \nonumber \\
 											&= \sum_{\substack{i \notin G_k \cup \{ k+1 \} \\ i \text{ not dangling}}} \sum_{j \in G_k} \frac{a_{ji}}{N_i} + \sum_{\substack{i \notin G_k \cup \{ k +1\} \\ i \text{ not dangling}}} \sum_{j \in G_k} \frac{1}{n}  \nonumber \\
 											&\tildeq  \sum_{\substack{i \notin G_k \cup \{ k +1\} \\ i \text{ not dangling}}} \sum_{j \in G_k} \frac{a_{ji}}{N_i} \nonumber \\
 											&\tildeq  \left( \sum_{\substack{i \notin G_k \\ i \text{ not dangling}}} \sum_{j \in G_k} \frac{a_{ji}}{N_i} \right) - \left( \sum_{j \in G_k} \frac{a_{jk}}{N_k} \right) \nonumber \\
 											&\tildeq - \frac{a_{jk}}{N_k}\sum_{j \in G_k}  \nonumber \\
||e^T P_{1:k, k+1:n}||_1			&= (n-k-1) \sum_{j \in G_k} P_{jk} = \frac{n-k-1}{N_k} \sum_{j \in G_k} a_{jk}  \nonumber
\end{align}
Now, why is expression \eqref{eq: efficient expression} more efficient? Because we keep an updated calculation of the terms $\delta_j$ in memory. After the first random walk initialization we compute $\delta_j$ for every $j$ in the sample. Then, whenever a node is added to the sample, they are updated. Namely, say that a node $s$ is 
added to $G_m$. Then for all the outgoing neighbours $j$ of $s$ already in $G_m$, we perform the update $\delta_j \leftarrow \delta_j - P_{js} = \delta_j - \frac{a_{js}}{N_j}$. \\
Summing up we get 
\begin{align*}
|| b_1^T U ||_1 	&= \frac{1}{N_k} \left(\alpha^2 \sum_{j \in G_k} a_{jk} \delta_{j} +  \frac{\alpha(1 - \alpha)(n-k-2)}{n} \sum_{j \in G_k} a_{jk}  \right)  \\
					&= \frac{\alpha}{N_k} \sum_{j \in G_k} a_{jk} \left(\frac{(1 - \alpha)(n - k - 2)}{n} + \alpha\delta_j \right)
\end{align*}
\end{itemize}
Notice that all the quantities here are expressed as a sum over all the nodes in $G_m$. However, the summands depend on the edges of the new nodes to be added, and can therefore be performed in $O(d_{in})$ or $O(d_{out})$. As opposed to $O(k)$, this is constant with respect to the sample size. \\
As a final remark, we would like to highlight the fact that it is much harder to find such a computational trick for the $L_2$ norm of the criterion vectors. This was instead possible for \textit{TCEC}, where they had a simpler expression that allowed derivations.

\section*{Appendix B. Details of the empirical implementation}
We set the leaderboard size to 100 for both \textit{TCEC} and \textit{TCPR}, and the $\alpha$ parameter for \textit{TCEC} to 0 for undirected networks, 0.5 for directed ones. The explorations are initialized with a random walk sampling of $1/5$ of the desired final sample size. The randomization level for neighbourhood exploration in set to $p=0.1$, meaning that $1/10$ of the possible nodes are explored, unless specified otherwise.

\section*{Appendix C. Extra plots for results on Pokec dataset}
\begin{figure}[htb]
\hspace{-0.8cm}\includegraphics[scale=0.2]{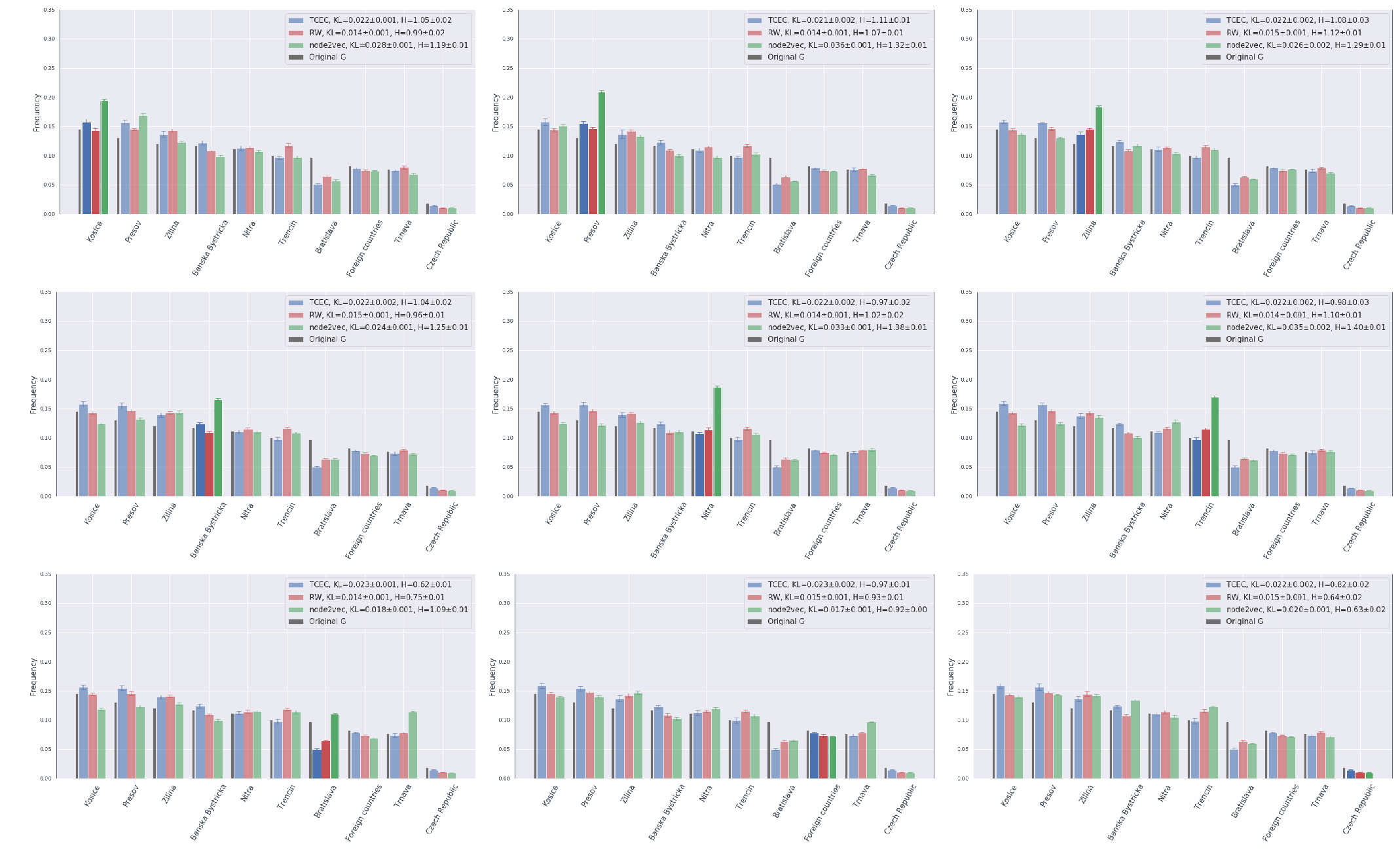}
\caption{\small Results of sampling on the Pokec dataset starting from nodes with different regions as attributes. The distribution on the whole network  is the black vertical line. Vertical lines on top of bars represent standard deviations across 10 runs of sampling. Numbers inside the legend are KL-divergence and entropy ratios between attribute distribution on the entire network and that inside the sample, as defined in the main text.}
\label{fig: appendix pokec}
\end{figure}

We include in Fig. \ref{fig: appendix pokec} the plots of the results on the Pokec dataset, similar to Fig. \ref{fig:pokec}, but where the seed point is chosen in each of the remaining regions. The bars relative to the initial region are bolded.

\newpage
\section*{Appendix D. Additional results on SBM for community sampling}
\begin{figure}[hpb]
\begin{centering}
\includegraphics[scale=0.25]{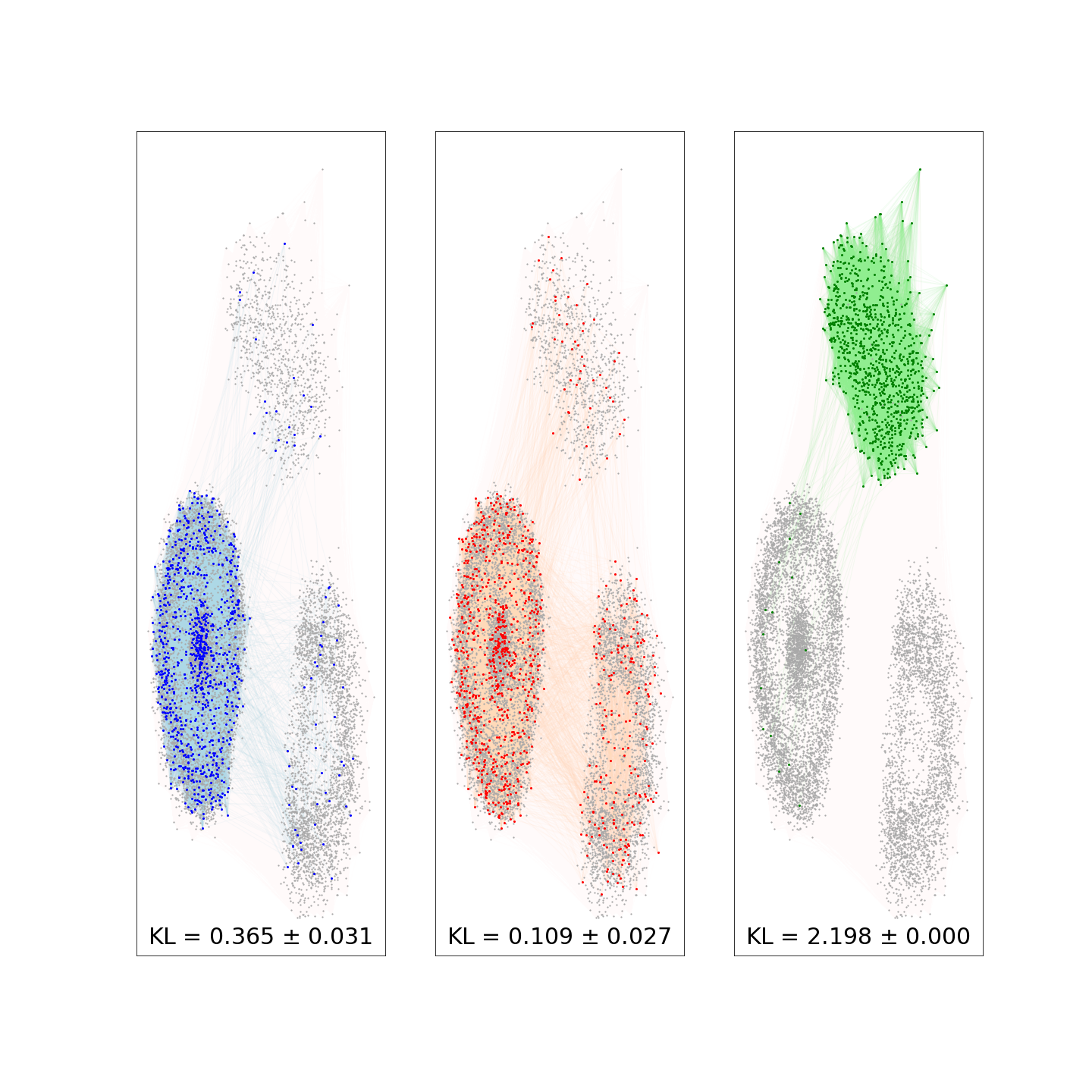}	
\end{centering}
\caption{\small Results of sampling on assortative SBM with different block sizes. 
 Sampled nodes and edges are colored in blue for \textit{TCEC} (left), red for uniform random walk  (center) and green for expansion sampling (right). The KL-divergence averages and standard deviations are computed over 10 different rounds of sampling $10\%$ of all the nodes.  }
\label{fig: SBM appendix}
\end{figure}

We include in Fig. \ref{fig: SBM appendix} the results for sampling on an assortative SBM structure with three unbalanced groups of sizes 1000, 3000 and 6000 nodes, within-block connection probability $p_{in} = 0.05$ and between-blocks connection probability $p_{out} = 0.005$

\end{document}